\documentclass[fleqn,10pt]{wlscirep}
\usepackage[utf8]{inputenc}
\usepackage[T1]{fontenc}
\title{Time \& its analytical definition: overcoming inconsistencies of previous results}

\author[1,*]{Umberto Lucia}
\affil[1]{Dipartimento Energia ``Galileo Ferraris'', Politecnico di Torino, Corso Duca degli Abruzzi 24, 10129 Torino, Italy (INFN-Sezione di Torino)}

\affil[*]{umberto.lucia@polito.it}

\usepackage{mathtools,amsfonts,amssymb,amsmath}
\usepackage{braket}
\usepackage{graphicx}
\usepackage{xcolor}
\usepackage{soul}
\setstcolor{red}
\usepackage{xcolor}
\usepackage{soul}\setstcolor{red}
\usepackage{cancel}
\begin{abstract}
Recently, a thermodynamic definition of time has been introduced. This definition is useful to find approach some open problems in physics. But, it was obtained by a phenomenological approach and a logical inconsistency appears in the definition. In particular, the definition was based on the ratio of two quantities, the entropy production and its rate, linked one another just by the definition of time. In this paper, this inconsistency is overcome, by using the second law of thermodynamics and Barbour's mathematical methods, obtaining an analytical result that brings to the same equation of the phenomenological method, but without any logical inconsistency.
\end{abstract}
\begin{document}
	
	\flushbottom
	\maketitle
	%
	%
	\thispagestyle{empty}
	
	
	\section*{Introduction}
    In Galilei's approach to the study of motion, time represents an absolute and fundamental quantity \cite{galileo}. Then, in Newton’s physics, time is a mathematical variable \cite{newton}. Any variation in physical quantity is always related to time. Isaac Newton considers time as only a mathematical entity, without any real or physical essence; simultaneity and durations of phenomena are absolute; any change in physical quantities occurs in a time interval between two points in an abstract mathematical space \cite{borghi-historioftime2016}. Movements and any other change in physical quantities occur in a space during a duration, that is, an extrinsic property measured by means of some motion of some part of a mechanical device \cite{borghi-clock2012,borghi-Hyp2013}.

On the contrary, Albert Einstein introduced a completely new viewpoint on the concept of duration in his Theory of Relativity \cite{einstein1905,einstein1917,einstein1920}, deriving the concept of time from the postulate of invariance of the speed of light, with the consequence that duration becomes a quantity dependent on the observer \cite{pauli-relativity}, i.e., the quantity measured by a clock synchronized concerning a given observer \cite{weinberg-gravitation}. In this context, time makes, together with the geometrical space, the spacetime, the coordinate system of our reality \cite{schrodingerspacetime1950}.

In the last decades, Loop Quantum Gravity has proposed a new viewpoint concerning General Relativity: no preferred clock time exists, while many clock times are measured by using different clocks, which obtain measures of time that undergo standard quantum
fluctuations like any other dynamical variable \cite{rovelli1991a,rovelli1991b,rovelli1993}.

In addition, thermodynamic approaches were developed about the measurable clock time concerning the perception of the human mind \cite{bjtime}: \textit{mind time} is explained as a sequence of images related to the inflow of stimuli \cite{bjvision} perceived by sensory organs. In this context, a constructal law approach \cite{bjtime,lorenzini-biserni} highlighted how changes in mental images decrease with age due to several physical features that change with age \cite{bjbigger} consequently to the evolution of flow brain architecture.

In all these approaches, time is used and defined in an operative way based on the concept of duration, but its analytical definition has not been introduced until 2009, when Julian Barbour used a mechanical approach for its definition \cite{barbour-natureoftime}. This approach is very interesting from an analytical viewpoint, but it does not consider the energy conservation of all its components, in particular heat.

To overcome this difficulty, an irreversible thermodynamic definition of time has been introduced  \cite{luciagrisolia-time-1,luciagrisolia-time-2,luciagrisolia-time-3} and applied to analyze some fundamental physical phenomena.
\cite{luciagrisolia-time-4,luciagrisolia-time-5,lucia-zeno}, starting from the analysis of irreversibility in open quantum systems \cite{lucia-atomicirr}. But this definition has been introduced only in a phenomenological way, by considering an intuitive approach to the effect of time on reality. This represents a limit of our approach because it results as a consequence of intuition but not a result of the fundamental laws of nature, i.e., Newton's laws, the quantum mechanics principles, the theory of relativity, and the thermodynamic laws.

Consequently, this paper aims to overcome this logical and theoretical inconsistency by introducing a physical-mathematical proof of our previous phenomenological definition, following the same analytical and rigorous physical-mathematical approach as Barbour but generalizing it by using the Laws of Thermodynamics to take into account any interaction (work in any form of energy variation and also heat).

	\section*{Results}
The result of this paper is to have analytically proven the results about time previously obtained by a phenomenological approach and published in Ref. \cite{luciagrisolia-time-2}. To do so, we follow the same rigorous mathematical method introduced by Barbour \cite{barbour-natureoftime}, but generalize it based on the laws of thermodynamics. Indeed, Barbour's approach is rigorous, but it considers only mechanical quantities, while in nature,  thermal and more general, not-only-mechanical quantities must be considered.

Time interval is proven not to exist independently from interaction; consequently, duration results from what the universe does, i.e., the interaction between electromagnetic waves (photons) and atomic or molecular electrons \cite{luciagrisolia-time-1,luciagrisolia-time-2,luciagrisolia-time-3}. This result is obtained by the second law of thermodynamics, concerning the entropy balance, which is related to the evolution of any system and to the footprint of any process.

Thus, time results strictly related to the evolution of the universe, in agreement with the General Theory of Relativity. Consequently, the analytical results allow us to obtain duration $\tau = t-t_0$, but if we consider its definition concerning a reference event $t_0$, we can use it as an analytical definition of time setting $t_0=0$ s. Thus, time results $t=\tau+t_0=\tau+0=\tau$.

	\section*{Discussion}
The definition of time represents oe of the open problems in physics. It is difficult to introduce a definition because all the events occur in time. 
 
 In Newtonian physics, time is a mathematical variable without any real base; any variation of physical quantities refers to time \cite{borghi-Hyp2013} and simultaneity and durations are considered absolute.

On the contrary, in Relativity, time is real quantity derived from the postulate of invariance of the speed of light, from which follows the relativity
 of simultaneity \cite{borghi-historioftime2016} and it flows at different rates for different observers. 
 
All physical systems evolve in the phase-space following the path of  entropic growth. Thus, the arrow of time emerges, linking thermodynamic irreversibility to quantum physics \cite{luciagrisolia-time-2}. 
 
 These considerations brings to a physical definition of time. Our previous definition was introduced in order to this requirement, but it was obtained only by a phenomenological approach based on the entropy ratio between entropy production and its rate, i.e., $\tau =\sigma/\Sigma$. The result highlighted that time is a footprint of irreversibility due to interaction between  atoms or molecules  and their environment. 
 
 Even if this result has been used in different context and application in order to allow us to find answers to some open questions in physics, the approach used was not so rigorous and analytical.
 
 So, some inconsistencies were present, just in the definition of ratio between two quantities linked bu time; indeed, entropy production rate is the first time differential $\Sigma = d\sigma/dt$ of the entropy production $\sigma$.
 
 In this paper, the inconsistency is overcome, by using the second law of thermodynamics and the Barbour's mathematical methods, obtaining a rigorous and analytical approach to obtain the same results of the phenomenological method, but without any inconsistency.

	
	\section*{Methods}
In this section we introduce the analytical proof of the results previously obtained by a phenomenological approach and published in Ref. \cite{luciagrisolia-time-2}.

To do so, we follows the same rigorous approach of Barbour \cite{barbour-natureoftime}, but in a more general way based on the laws of thermodynamics. Indeed, the Barbour's approach is rigorous, but it considers only mechanical quantities, while in Nature mechanical and thermal, and more general not-only-mechanical quantities, must be considered. Moreover, time is strictly related to the evolution of any process, so we must consider the thermodynamic quantity which allows us to describe the evolution of any real process, the entropy.

So, we consider the Second Law of Thermodynamics \cite{bejan-advengtherm}:
\begin{equation}
	\frac{dS}{dt}=\frac{\phi}{T} + \Sigma
\end{equation}
where $S$ is the entropy [J K$^{-1}$], $\phi$ is the heat power [W], $T$ is the temperature [K], $\Sigma = d\sigma/dt$ is the entropy production rate [W K$^{-1}$], with $\sigma$ entropy production [J K$^{-1}$] and $t$ time [s]. For a stationary system $dS/dt =0$, so it follows:
\begin{equation}
	\frac{\phi}{T} + \frac{d\sigma}{dt}=0
\end{equation}
obtaining:
\begin{equation}
	dt = -\frac{d\sigma}{\phi/T}
\end{equation}
from which a time interval, $\tau$ [s], can be defined as:
\begin{equation}
	\tau = -\frac{T\,\sigma}{\phi}
\end{equation}

Recently, a thermodynamic approach \cite{lucia-atomicirr,lucia-zeno} to irreversibility in quantum systems has been developed based on the experimental evidence that irreversibility is caused by the continuous interaction between the environmental electromagnetic waves and the matter \cite{exp1,exp2,exp3,exp4}. But, radiative processes in matter and its link to entropy variation are well known processes \cite{planck59,heitler10,jammer66,surdin66,rueda74,fonseca09,boyer18,berettaetal2015}.

So, considering that the temperature in this analysis is the environmental temperature $T_0$, as a consequence of the Gouy-Stodola theorem, the entropy production rate $\Sigma$ results \cite{berettaetal2015}:
\begin{equation} \label{entropyratevar}
	\phi= \frac{A}{2}\varepsilon_0 c E_{el}^2 + \frac{A}{2\mu_0} c B_m^2
\end{equation}
where  $E_{el}$ is the electric field, $B_m$ is the magnetic field, $c = 299 792 458$ m s$^{-1}$ is the speed of light, $\varepsilon_0 = 8.8541878128(13)\times 10^{-12}$ F m$^{-1}$  is the electric permittivity in vacuum and $\mu_0  4\pi\times 10^{-7}$ H m$^{-1}$ is the  magnetic permeability in vacuum, $A$ is the area of the border of the thermodynamic control volume, and $T_0$ is the environmental temperature, while the entropy production has been evaluated with respect to the semi-classical analysis of the photon-bound electron interaction \cite{lucia-atomicirr}:
\begin{equation}
	T_0 \sigma= \frac{m_e}{M}\, E_\gamma
\end{equation}
where $T_0$ is the environmental temperature, and $E_\gamma$ is the energy of the incoming photon. So, the time interval has been evaluated by means of measurable physical quantities as follows:
\begin{equation}
	\tau =\frac{2\,m_e}{M\,c\,A}\,\frac{E_\gamma}{\varepsilon_0\, E_{el}^2 + \mu_0^{-1} \, B_m^2}
\end{equation}

	
	\bibliography{Biblio}
	
	
	
	
	
	
	
	\section*{Additional information}
	
	 \textbf{The author declares no competing interests}.
	
	\section*{Data availability statement}
	 All data generated or analysed during this study are included in this published article.


%
	
\end{document}